\documentclass[final,english]{bullsrsl}[2022/06/15]

\usepackage[latin1]{inputenc}
\usepackage[T1]{fontenc}

\usepackage{natbib} 
\usepackage{graphicx}

\begin{document}
\title{Investigation of orbital period changes in 9 contact binaries}

\author[affil={1}, corresponding]{Yogesh Chandra}{Joshi}
\author[affil={1,2}]{Alaxendra}{Panchal}
\affiliation[1]{Aryabhatta Research Institute of Observational Sciences (ARIES), Nainital}
\affiliation[2]{Department of Physics, DDU Gorakhpur University, Gorakhpur, India}
\correspondance{yogesh@aries.res.in}
\date{11 May 2023}

\maketitle


%

\begin{abstract}
We present the results for orbital period analysis of 9 contact binaries (CBs). The photometric data analysed in this work is collected using ARIES 1-m and 1.3-m telescopes as well as many ground and space-based photometric surveys. The precise orbital periods of the binary systems are studied using the long time-baseline of data acquired over the last 12-15 years. The changes in the times of minimum brightness are calculated using (O-C) diagram. Out of these 9 CBs, four systems show no change in the orbital period with time while the remaining five systems show non-linear (O-C) variations with time. We derive mass transfer rates for these five CBs which suggests mass is being transferred from secondary to primary components in three systems while it is from primary to secondary components in the other two systems.
\end{abstract}


%
\section{Introduction}
Eclipsing binaries (EBs) are a type of binary system in which flux variation occurs whenever one component eclipses another component depending on their specific alignment towards the observer. EBs are popular among celestial sources as they are used for precise  mass, radius and distance measurements \citep{2016RAA....16...63J, 2017RAA....17..115J}. The number of EBs has increased tremendously due to recent ground and space-based missions. A large fraction of the EBs observed in these missions are categorized as contact binaries (CBs). The CBs are easier to detect as they show continuous flux variation and have period less than one day. The CBs cover more than 50\% of the overall EB population detected by missions like Catalina Real-Time Transient Survey (CRTS), the All Sky Automated Survey for SuperNovae (ASAS-SN), Kepler, etc. Apart from being a well-known parameter and distance measurement tool, CBs provide a way to study the interaction between the binary components \citep{1962ApJ...136..312K}. Furthermore, understanding a relationship between the mass ratio and the orbital period of CBs as well as orbital period change over time has been an interesting topic for a long time \citep{2022AJ....164..202L, 2023RAA....23d5017C}. An orbital period change can be associated with mass transfer, angular momentum loss due to magnetic stellar winds, or a light-time effect because of the presence of a third component in the system. Long-term period analysis of such systems therefore can be used to detect period variations and study associated physical mechanisms \citep{2021AJ....161..221P, 2022ApJ...927...12P}. The present study shows results from our orbital period analysis for 9 CBs. The sources are selected from the Catalina Surveys Periodic Variable Star Catalog. All these systems are classified as contact binaries and have an orbital period less than 15 hrs.

\section{Observations}
To study orbital period behaviour on a large time scale, we searched these targets in all previous photometric archives. For the ground-based photometric observations of these targets, we used the 1.3\,m Devasthal Fast Optical Telescope (DFOT) and the 1.04\,m Sampurnanand Telescope (ST), operated by the Aryabhatta Research Institute of Observational Sciences (ARIES). The DFOT employs a $2k\times2k$ CCD detector having a field of view of $\sim 18^{'} \times 18^{'}$ \citep{2022JAI....1140004J}. The ST is equipped with a 4k$\times$4k CCD with a FoV of $\sim$ 15$^{'}\times$15$^{'}$. All the photometric data was processed with the IRAF software while instrumental magnitudes of stellar sources were computed using the DAOPHOT. The differential light curves were generated after choosing suitable comparison stars close to the target stars in the field. Some sources were observed by SuperWASP which is an wide-angle photometric survey covering almost 500 $deg^2$ in the northern hemisphere. The archive contains time series data of almost 18 million sources observed during the survey. Apart from SuperWASP survey, we also used data from CRTS, Palomar Transient Factory (PTF), Zwicky Transient Facility (ZTF), and ASAS-SN for some sources.

For space-based data, we used archival data provided by the Transiting Exoplanet Survey Satellite (TESS) survey that uses four cameras with a FoV of 24$^{\circ}\times$24$^{\circ}$ each as back-end instruments. The TESS data is publicly available at the Barbara A. Mikulski Archive for Space Telescopes (MAST) portal \footnote{https://mast.stsci.edu}.

\section{Methodology}
We analyzed the time series for these objects to detect any secular changes in their orbital period. The precise period of these 9 binary systems was determined using the Period04 software that employs a discrete Fourier Transform Algorithm. The orbital period variation with time is investigated by studying the variations in time of minima (TOMs) at different epochs. For this purpose, we are required to extract the observed and calculated (O-C) TOM for each EB at different epochs ($E$). Here, we only selected the good quality light curves having less scattering and better phase coverage around the minima. The TOMs for the primary or secondary eclipse were estimated with parabola fitting to the data around primary or secondary minima. The shape of (O-C) plot reveals the behaviour of the evolution of the orbital period of the system.
$$ JD_{cal} = JD_{o} + Period \times E $$
$$ (O-C) = JD_{obs} - JD_{cal}  $$
The (O-C) analysis of an EB primarily involves four steps which are highlighted below:
\begin{enumerate}
\item Observe EBs near primary or secondary eclipse.
\item Fit parabola in minimum brightness region to get observed TOM (JD$_{obs}$).
\item Determine calculated TOM (JD$_{cal}$) using average orbital period.
\item Plot (O-C) with respect to different orbital cycles and fit the polynomial.
\end{enumerate}%
\begin{figure}
\centering
\vspace{-0.5cm}
\includegraphics[width=5.2cm, height=4.5cm]{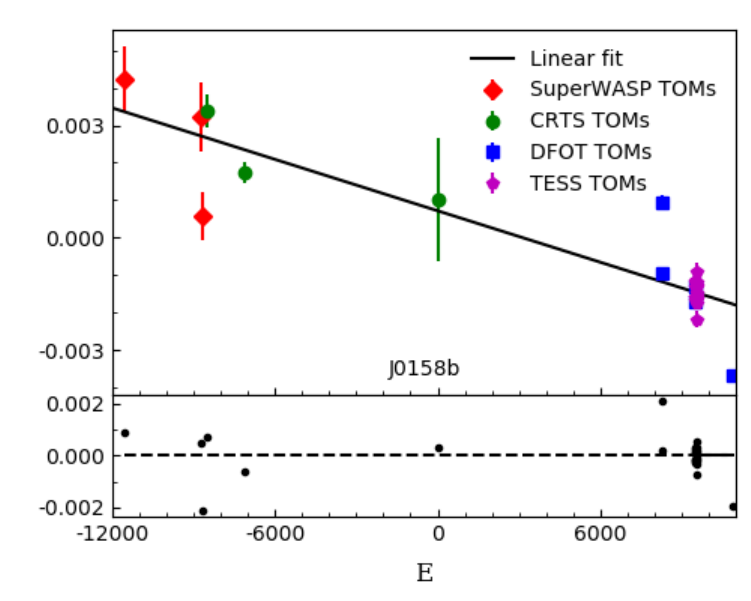}
\includegraphics[width=5.2cm, height=4.5cm]{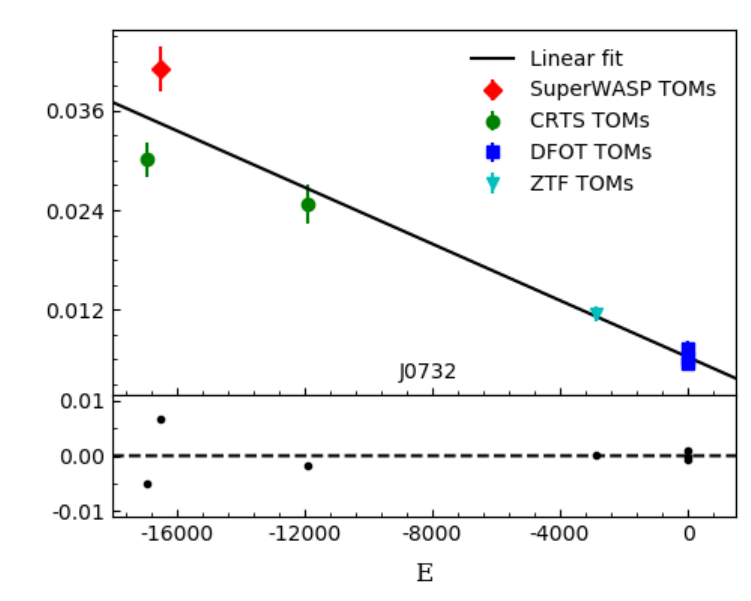}
\includegraphics[width=5.2cm, height=4.5cm]{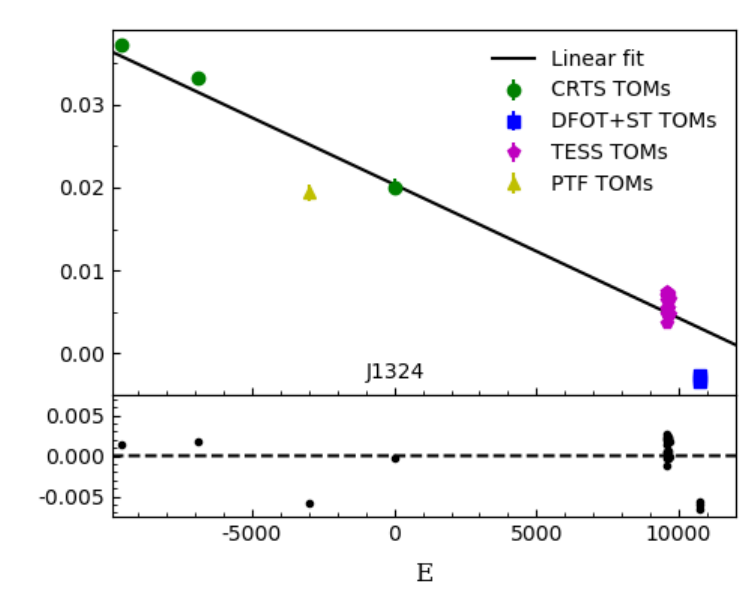}
\includegraphics[width=5.2cm, height=4.5cm]{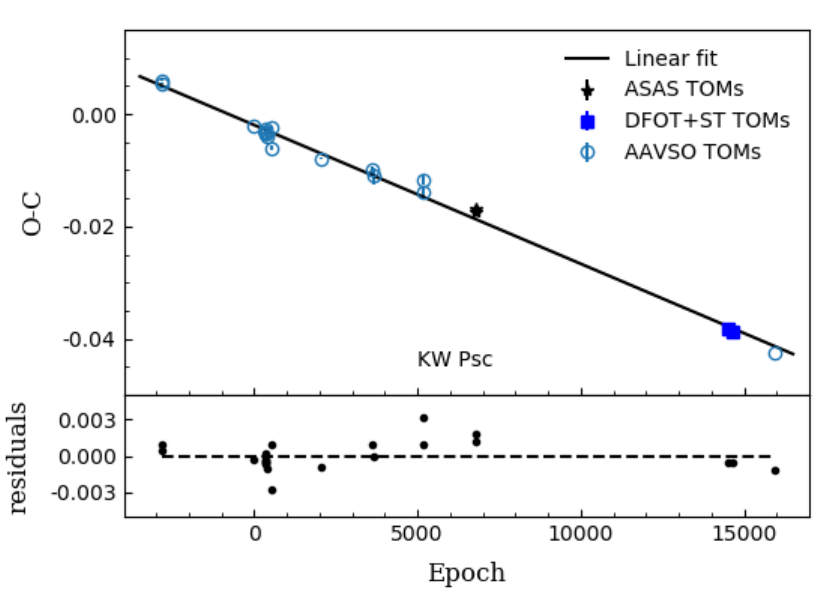}
\includegraphics[width=5.2cm, height=4.5cm]{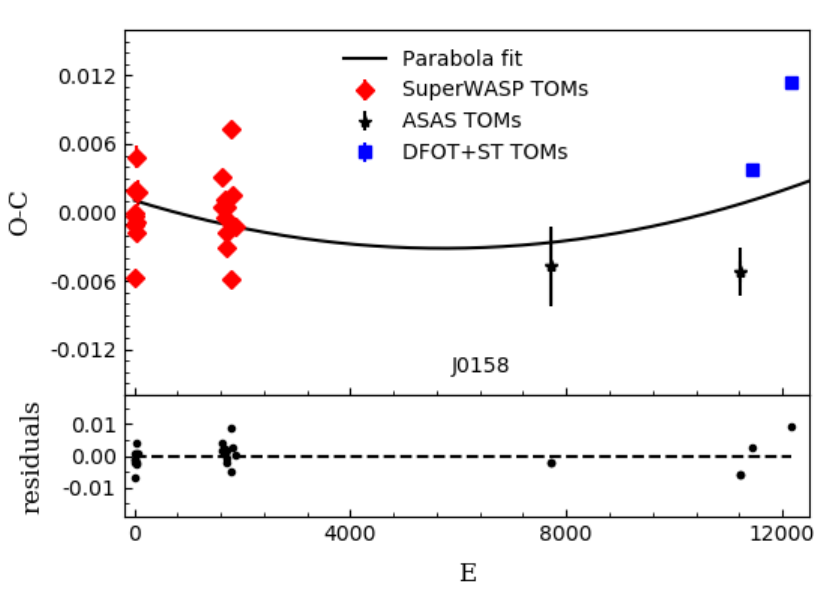}
\includegraphics[width=5.2cm, height=4.5cm]{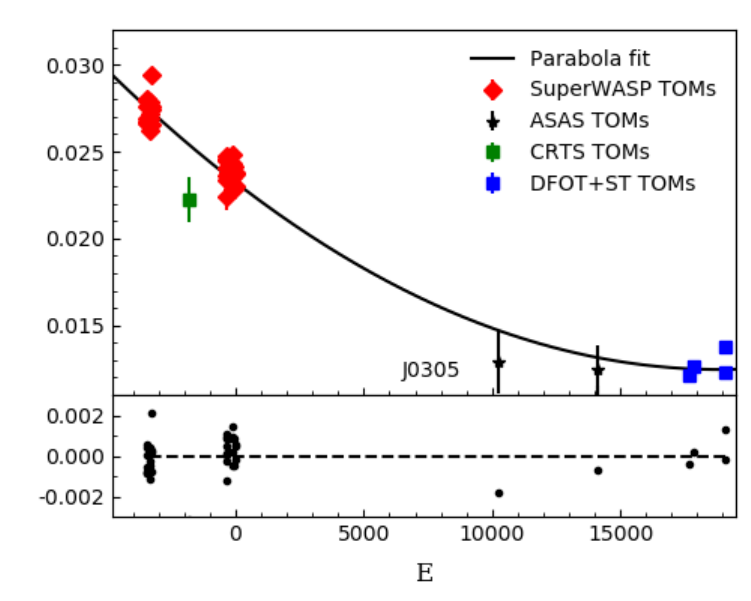}
\includegraphics[width=5.2cm, height=4.5cm]{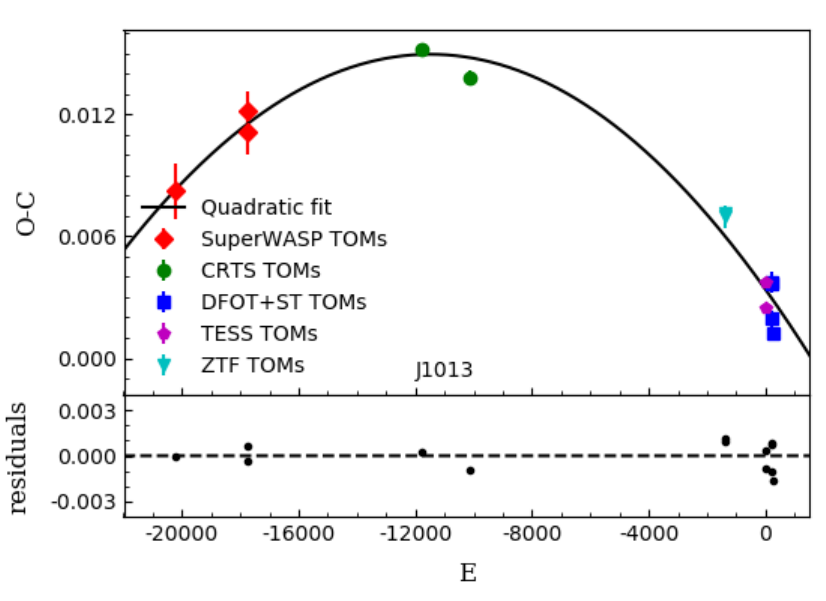}
\includegraphics[width=5.2cm, height=4.5cm]{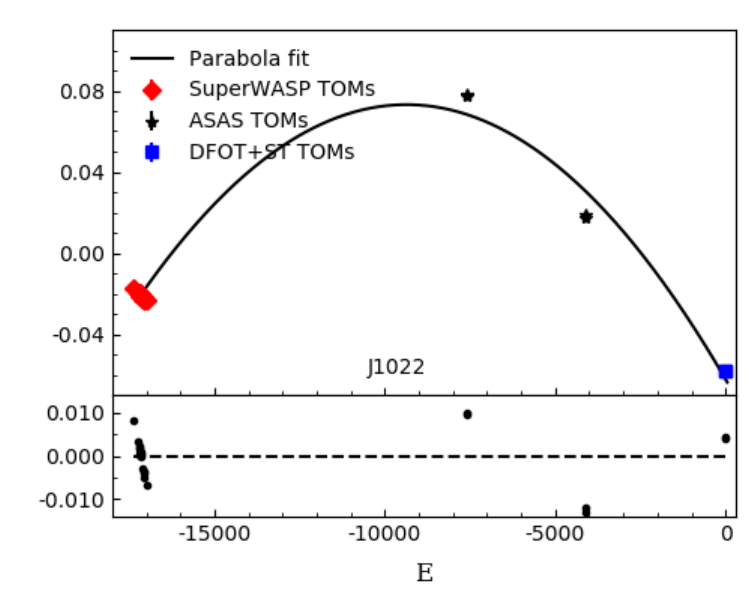}
\includegraphics[width=5.2cm, height=4.5cm]{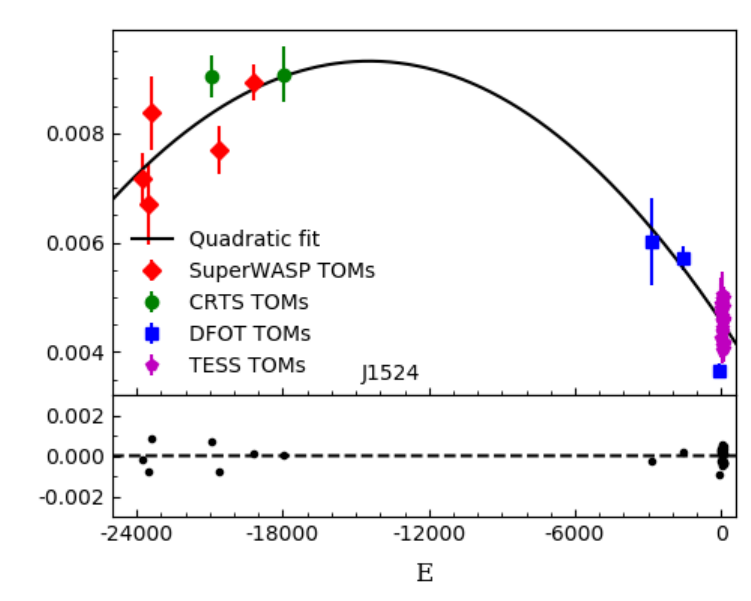}
\caption{The (O-C) diagrams of the 9 CBs with linear and quadratic fits. The x-axis corresponds to the orbital cycle number and y-axis represents (O-C) difference of TOMs in days. The residual of the fit is shown in the lower panel in each plot. }
\label{fig01}
\end{figure}
%
%
\section{Analysis}
In our orbital period analysis, we detected no period variation in four systems while the remaining five systems show parabolic variation in their (O-C) diagram which indicates a non-variable period. The (O-C) diagram for each system is illustrated in Figure~\ref{fig01}. We also plot the residuals of the fit in the bottom panel of each plot, indicating no particular pattern with epoch suggesting the robustness of the fit to the observed data points. However, we can see some scatter in the residuals for some systems, which may be caused by the asymmetry and the variation of the light curve due to the presence of the dark and/or hotspots on the surface of one or both the components of the binary systems. These figures show (O-C) diagrams following the polynomials given below: 
$$O-C (J0158b) = 0.0007 (\pm0.0002) - 2.2901  (\pm0.2148)  \times 10^{-7}\times E $$
$$O-C (J0732)  = 0.006  (\pm0.002)  - 1.7065  (\pm0.1981)  \times 10^{-6}\times E $$
$$O-C (J1324)  = 0.0204 (\pm0.0009) - 1.611598(\pm0.103038)\times 10^{-6}\times E $$
$$O-C (KWPsc)  =-0.00195(\pm0.00030)- 2.47017 (\pm0.04911) \times E$$
$$O-C(J0158)   =-0.0010 (\pm0.0011) - 1.467  (\pm0.906)  \times 10^{-6}E + 1.283 (\pm0.772) \times 10^{-10} \times E^{2}$$
$$O-C(J0305)   =-0.0005 (\pm0.0002) - 1.799  (\pm0.730)  \times 10^{-7}E + 3.004 (\pm0.472) \times 10^{-11} \times E^{2}$$
$$O-C(J1013)   =+0.0034 (\pm0.0004) - 2.011  (\pm0.156)  \times 10^{-6}E - 8.746 (\pm0.852) \times 10^{-11} \times E^{2}$$
$$O-C(J1022)   =-0.0003 (\pm0.0004) + 1.901  (\pm1.339)  \times 10^{-7}E - 7.825 (\pm0.678) \times 10^{-11} \times E^{2}$$
$$O-C(J1524)   =-0.0046 (\pm0.0001) - 6.582  (\pm0.701)  \times 10^{-7}E - 2.277 (\pm0.319) \times 10^{-11} \times E^{2}$$
%
\subsection{Orbital Period Variation}
%
As seen in Figure~\ref{fig01}, four systems show linearity in the (O-C) diagram which suggests that periods did not change in the last 12-15 years of our observing span. However, non-linear variation is quite apparent in the (O-C) diagram of the remaining five systems. The (O-C) variations for these five systems were fitted with the quadratic polynomials. The rate of period change is determined for these five systems using the quadratic term (c) with the help of following relation:
$$ dP/dt = 2c/P $$
The period and the estimated rates of period change ($dP/dt$) for these five systems are given in Table~\ref{tab01}.

\begin{table}[!ht]   
\centering
\caption{Period ($P$), primary mass ($M_{1}$), mass-ratio ($q$) and rate of period change ($dP/dt$) for five systems.}     
\begin{tabular}{|c |c |c |c |c |}    
\hline  
Source      &     P    &       M$_{1}$      &     q            & dP/dt      \\
            &   (days) &   (M$_{\odot}$)    &                  &(days year$^{-1}$) \\
\hline
 & & & & \\
J0158       & 0.455333 & 1.262 ($\pm$0.171) & 0.67 ($\pm$0.12) & +2.057 ($\pm$1.238) $\times 10^{-7}$    \\
J0305       & 0.246983 & 0.927 ($\pm$0.085) & 0.31 ($\pm$0.01) & +8.879 ($\pm$1.395) $\times 10^{-8}$    \\
J1013       & 0.250205 & 0.810 ($\pm$0.060) & 0.45 ($\pm$0.01) & -2.552 ($\pm$0.249) $\times 10^{-7}$    \\
J1022       & 0.258484 & 0.313 ($\pm$0.027) & 3.23 ($\pm$0.14) & -2.210 ($\pm$0.191) $\times 10^{-7}$    \\
J1524       & 0.244759 & 1.040 ($\pm$0.060) & 0.39 ($\pm$0.01) & -6.792 ($\pm$0.952) $\times 10^{-8}$    \\
 & & & & \\
\hline                  
\end{tabular}  
\label{tab01}     
\vspace{1ex}
\end{table}

In the contact stage, a positive rate of period change suggests the mass transfer is going on from the secondary to the primary components for systems with $q<1$, while the mass transfer is from primary to secondary for the systems showing a negative rate of period variations for systems with $q<1$.
%
\subsection{Mass Transfer Rate}
%
The period analysis of the 9 CBs reveals that five of them exhibit changes in the orbital period. Although different processes can affect the orbital period in the contact binaries, mass-loss or mass transfer between the components is the prime reason in most of the cases as the effect of processes like Gravitational wave radiation or magnetic braking are usually very small. The mass-ratio (q) of a binary system ($q = M_2/M_1$), is very critical to study the evolution of binary components. Using the $q$ and individual component masses information for these systems from \cite{2021AJ....161..221P} and \cite{2022ApJ...927...12P}, we calculated the mass-transfer rates for variable period systems with the help of mass-transfer relation given by \cite{1958BAN....14..131K} as following.
\begin{equation}
\label{matr}
\frac{1}{M_{1}}\frac{dM_{1}}{dt}=\frac{q}{3(1-q)}\frac{1}{P}\frac{dp}{dt}
\end{equation}
The change in the mass of the primary component were estimated as given below. The q and $M_{1}$ for each of these five systems are given in Table~\ref{tab01}.
$$dM1 /dt (J0158) =  3.858 (\pm3.169) \times 10^{-7} M_\odot year^{-1}$$
$$dM1 /dt (J0305) =  4.991 (\pm0.937) \times 10^{-8} M_\odot year^{-1}$$
$$dM1 /dt (J1013) = -2.253 (\pm0.290) \times 10^{-7} M_\odot year^{-1}$$
$$dM1 /dt (J1022) =  1.292 (\pm0.159) \times 10^{-7} M_\odot year^{-1}$$
$$dM1 /dt (J1524) = -6.150 (\pm0.967) \times 10^{-8} M_\odot year^{-1}$$
Here, a negative $dM_{1} /dt$ corresponds to mass transfer taking place from the primary component to the secondary component in the binary system.
\section{Results and Discussion}
The study of EBs is important to understand star formation and evolution. As most of the binaries are expected to be a part of multiple systems with faint or wide orbit extra components, orbital period analysis of EBs can give clues about multiplicity among stellar systems. Although complete characterization requires photometric and spectroscopic observations, photometric data alone is useful to understand orbital period evolution by studying their period variation with time.

In this work, we studied the orbital period behaviour of 9 CBs using the photometric time series data from our ground-based observations and previously available ground and space-based surveys to detect possible changes in their orbital period. Our (O-C) analysis of these systems revealed a change in the orbital period for five systems but no such variation was noticed in the remaining systems though our interpretation is based on the TOM information collected during the last 12-15 years only. The mass loss is often interpreted as a reason for change in their periods hence period change rates are converted into mass-transfer rates between components. Since both components are close to each other in CBs, the interaction between the two components is quite common in these systems. This can result in a period change through mass transfer or mass loss among components. The presence of additional companion or long-term cyclic magnetic activity is also prevalent for contact binaries which can cause cyclic variation in the (O-C) diagram.

Such types of studies are helpful to understand the physical interaction between binary components, mass transfer processes, and magnetic activities and detect extra components in the system. Since period variations can contribute to the direct determination of mass transfer and angular momentum loss, further long-term observations for these systems will help to understand the structure and evolutionary state of these binary systems as well as to detect any long-term variation in the binary systems or possibility of any other component in the systems.
%
\begin{furtherinformation}

\begin{orcids}   
\orcid{0000-0001-8657-1573}{Yogesh}{Joshi}
\end{orcids}

\begin{authorcontributions}
The formulation or evolution of present research topic is envisaged by the lead author of the paper including the major writing of the draft while data acquisition and analysis has been done by Alaxendra Panchal. 
\end{authorcontributions}

\begin{conflictsofinterest}
The authors declare no conflict of interest.
\end{conflictsofinterest}

\end{furtherinformation}

\bibliographystyle{bullsrsl-en}

\bibliography{reference}

\end{document}